\UseRawInputEncoding

\documentclass[a4paper,scale=0.8,onecolumn,nobibnotes,nofootinbib]{revtex4}
\usepackage{enumitem}
\usepackage{amsmath,amssymb}
\usepackage{graphicx} 
\usepackage{epstopdf}
\usepackage{hyperref}
\usepackage{geometry}
\usepackage{subfigure}
\usepackage[section]{placeins}
\usepackage{color}

\hypersetup{
    colorlinks=true,
    citecolor=blue,
    linkcolor=red,
    filecolor=magenta,
    urlcolor=cyan,}

\newcommand{\be}{\begin{equation}}
\newcommand{\ee}{\end{equation}}

\begin{document}

\title{The neutrino behavior in the Einstein$\text{-}$Hilbert$\text{-}$Bumblebee gravity around global monopole field}

\author{ B. Q. Wang \footnote {Corresponding author:wangbingqian@gzy.edu.cn}$^{1}$}
\author{ S. R. Wu \footnote {wushurui@gyu.edu.cn}$^{2}$}
\author{ Z. W. Long  \footnote {zwlong@gzu.edu.cn}$^{3}$}

\affiliation{$^{1}$ College Pharmacy, Guizhou University of Traditional Chinese Medicine, Guiyang 550025, China.\\ $^2$ School of Science, Guiyang University, Guiyang, 550025, China. \\ $^3$  College of Physics, Guizhou University, Guiyang, 550025, China. }
\date{\today}

\begin{abstract}

In this work, we study the neutrino oscillation phase propagating along radial and non-radial paths in the Einstein$\text{-}$Hilbert$\text{-}$Bumblebee (EHB) gravity around global monopole field, by using the quantummechanical treatment, the expression of the corrected neutrino oscillation probability is obtained. Our results show that neutrino oscillation in the EHB gravity around global monopole field is different from that in Schwarzschild black hole, the Lorentz-violating parameter $a$ mainly affects the peak of the oscillation probability, and the global monopole $\overline { \mu }$ and the lightest neutrino mass both affect the frequency of the oscillations of probabilities, which means that studying the neutrino oscillation phenomenon in curved spacetime may not only become a new technology for probing the properties of compact celestial bodies, but also help deepen our understanding of neutrinos.

\end{abstract}

\maketitle
~~~~~~~~~\textsl{Keywords:}  neutrino oscillation, gravitational lensing, Einstein$\text{-}$Hilbert$\text{-}$Bumblebee gravity.
\section{Introduction}

In the standard model of particle physics, the neutrino is an elementary particle, often represented by the symbol $\upsilon$, moreover, as a species of lepton, neutrinos do not participate in strong interactions and have a spin of $\hbar/2$. But unlike other leptons, neutrinos can only participate in weak and gravitational interactions because their electric neutrality does not participate in electromagnetic interactions. At present, the study of the fundamental properties of neutrinos mainly focuses on the phenomenon of neutrino oscillation. In the atmospheric neutrino oscillation experiment, early IMB, MACRO and the Super-Kamiokande probe in Japan observed a deviation in the ratio of muon neutrinos to electroneutrinos emitted from the atmosphere \cite{retMLO}.

The theoretical prediction of the observed number of neutrinos reveals a constant value, which does not vary with the zenith angle. However, according to the result from the Super-Kamiokande probe in 1998, muon neutrinos coming in from below the detector (produced on the other side of the Earth) were observed at half the number of muon neutrinos coming in from above the detector \cite{retYFU}. Subsequently, this result can be interpreted as a shift or oscillation of neutrinos to other types of undetected neutrinos, which is now known as neutrino oscillation. This new finding indicates that the standard model is incomplete and needs further extension, because it proves that neutrinos have a finite mass, and when neutrinos oscillate between three flavour eigenstates, each has its own rest mass. Further research in 2004 reported that the event rate is a function of length divided by energy and has a sinusoidal correspondence, verifying the phenomenon of neutrino oscillation. However, since the mass of neutrinos is too small, it is extremely difficult to measure the absolute mass of neutrinos in experiments. For example, M.~Tanabashi \textit{et al.} \cite{retMTA} based on oscillations have measured ${\Delta m}_{21}^2$ and ${\Delta m}_{31}^2$, they leaves two possibilities: (1) $m_1<m_2<m_3$, the normal ordering, (2) $m_3<m_1<m_2$, the inverted ordering, here $m_1$,  $m_2$, $m_3$ represent masses corresponding to the neutrino mass eigenstates. Therefore, physicists try to find a solution in theory, for example, there have been several studies on the effects of curved spacetime in terms of neutrino oscillations \cite{retCYC,retNFO,retJGP,retGLA,retSIG,retSCH,retLBU,retKBO,retACA,retHSW}, especially in \cite{retHSW}, it shows that unlike in the flat spacetime neutrino, neutrino flavor transition probability in the weak gravity neutrino lensing contains the information about the absolute neutrino mass and their hierarchy, this clearly provides an inspiration for future experiments to measure the absolute mass of neutrinos.

In this work, we study the neutrino behavior in the EHB gravity around global monopole field. In this line of thought, considering the existence of lensing around a massive astrophysical object, the oscillation probability at a particular point is calculated by analysing all possible trajectories of neutrinos which get focused at that point due to the lensing phenomenon \cite{retChakrabarty}. General relativity and the standard model of particle physics are two important theoretical cornerstones of modern physics, and unifying these two theories would give us a deeper understanding of nature. The theory of quantum gravity is an important product in the process of seeking the grand unification of theories. When it comes to the numerous modified gravitational theories, a possible extension of general relativity is the theory of gravity with Lorentz symmetry breaking \cite{retJacobson,retHorava}, which is related to the view that the quantum gravity signals are likely to be detected at low energy scales \cite{retCASA}, and another compelling point is that this work \cite{retCASA} opens up the scope of the static and spherically symmetric black hole solutions in bumblebee gravity, subsequently this research is widely applied to various scenarios, including the scalar quasibound states in Einstein-Maxwell-Bumblebee black holes \cite{retSenjaya:2025cgk}, the field perturbations and observables in a charged black hole within bumblebee gravity \cite{retSingh:2025fuv}, the energy extraction from a rotating black hole via magnetic reconnection \cite{retYuChih:2025hsg}, the electrodynamic black holes in Bumblebee gravity with Lorentz Violation and shadow phenomenology \cite{retSekhmani:2025gvv}, the applications of gauss-bonnet theorem \cite{retOvgun:2018ran}, the ergosphere dynamics and rotational energy extraction \cite{Hassanabadi:2025ect} and the quasinormal modes \cite{retDeng:2025uvp}. As a pillar of modern physics, Lorentz symmetry may not be valid at the Planck scale \cite{retMattingly}, therefore testing possible Lorentz symmetry breaking signatures in the gravitational sector is becoming a hot research direction, especially in terms of how this mechanism affects the thermodynamics of black holes \cite{retFilho:2021rin,retAnacleto:2018wlj,retFilho:2020cgk,retFilho:2023ydb}. The bumblebee gravity is one of this Lorentz violation gravitational theories, thus it has been received a lot of attention \cite{retShi:2025rfq,retShi:2025plr}. On the other side, spontaneously broken global symmetries give rise to a divergence of the global monopole energy in terms of the monopole case, a common approach dealing with this problem is to introduce the self gravity of global monopole. Global monopoles are point-like topological defects, which are assume that emerge during symmetry-breaking phase transitions in the early universe \cite{retBarriola:1989hx}, especially involving scalar fields, they undergo a process from the global O(3) symmetry into U(1) symmetry during phase transitions in the early universe \cite{retVilenkin:1981kz,retVilenkin:1984ib}. Recently, various scenarios are discussed in terms of global monopoles, including the phantom global monopole \cite{retNooriGashti:2025vyg}, the ray geodesic and wave propagation in lorentz-violating wormholes \cite{retAhmed:2025yyb}, the global monopole induced wormholes in power-law gravity \cite{retYousaf:2025jqx}, the black holes surrounded by a Murnaghan-fluid scalar gas \cite{retSekhmani:2025yjf} and the effect of global monopole on Schwarzschild spacetime \cite{retDadhich}.

Based on the above considerations, in this manuscript we focus on the neutrino behavior in the EHB gravity around global monopole field aiming to study the effect of the weak gravitational lensing on the neutrino oscillation. The article is organised as follows: a brief review of Schwarzschild-like black hole with a topological defect in bumblebee gravity and the theoretical background and assumptions of neutrino oscillation are provided in sections \ref{sec2} and \ref{sec3}, respectively. In section \ref{sec4}, we discuss the neutrino oscillation phase in the EHB gravity around global monopole field and in section \ref{sec5} the neutrino oscillation probability in the EHB gravity around global monopole field is studied, the numerical results of two flavor case is discussed in section \ref{sec6}, finally we summarize and comment on our results in section \ref{sec7}.

\section{Schwarzschild-like black hole with a topological defect in bumblebee gravity}\label{sec2}

 The Schwarzschild-like solution in the EHB gravity around global monopole field formulated within the metric formalism
 \begin{equation}
d s ^ { 2 } = - \left( 1 - \overline { \mu } - \frac { 2 M } { r } \right) d t ^ { 2 } + ( 1 + a ) \left( 1 - \overline { \mu } - \frac { 2 M } { r } \right) ^ { - 1 } d r ^ { 2 } + r ^ { 2 } d \theta ^ { 2 }  + r ^ { 2 } ( \sin \theta ) ^ { 2 } d \phi ^ { 2 },
 \label{eq1}   \end{equation}
where $M$, $a$ and $\overline { \mu }$ represent the usual geometrical mass, the Lorentz-violating parameter and the global monopole term respectively. By introducing the coordinate transformation, the event horizon of the black hole locates at $g_{tt} (r_h )=0$ so that $ r_h= 2M$, which does not depend on $a$ and $\overline { \mu }$, this implies that bumblebee field and global monopole field has no effect on event horizon \cite{retGullu:2020qzu}. On the other hand, there exists a physical singularity at $r = 0$ because of the divergence of the Kretschmann invariant. Moreover, the Kretschmann scalar is different than the Schwarzschild case, which means that it is impossible to recover Eq. \eqref{eq1} to Schwarzschild spacetime through a coordinate transformation, because this formalism is a real solution containing the global monopole term and the Lorentz-violating corrections. Therefore, in this paper we only consider the gravitational field in the range of radial coordinates $r\in\left(2M,\infty\right)$.

Here for simplicity, we shall name the metric coefficients as follow
\begin{equation}
\begin{aligned}
 &g_{tt}=-A,   g_{rr}=\frac{1+a}{A}, \\  &g_{\theta\theta}=C,   g_{\phi\phi}=D,   \label{eq2}
\end{aligned}
\end{equation}
where $A=\left(1-\overline { \mu }-\frac{2M}{r}\right),C=r^2,D=r^2\left(\sin{\theta}\right)^2$ and the corresponding contravariant component of the metric can be given by the formula $g_{\mu\nu}g^{\mu\nu}=1$.

\section{Theoretical background and assumptions of neutrino oscillation}\label{sec3}

Neutrino oscillations are rooted in the fact that their weak interaction flavour eigenstates $\lvert\ \upsilon_\alpha\rangle$ are different from their propagation mass eigenstates $\lvert\ \upsilon_i\rangle$, the relationship between these two eigenstates can be described by
\begin{equation}
\left\lvert\left.\ \upsilon_\alpha\right\rangle\right.=\sum_{i}{U_{\alpha i}^\ast\left\lvert\left.\ \upsilon_i\right\rangle\right.} ,   \label{eq3} \end{equation}
where $i=1,2,3$ and $U_{\alpha i}^\ast$ is a 3x3 neutrino flavor mixing unitary matrix \cite{retMaki}.
Here we assume that the neutrino wave-function have a plane-wave wavefunction \cite{retSwami} and it propagates in vacuum from source $S(t_S, \boldsymbol{x}_S)$ to detector $D(t_D,\boldsymbol{x}_D)$, thus
during the process of propagation, the phase accumulation of $\lvert\ \upsilon_i\rangle$ can be described by $\left\lvert\upsilon_i\left(t_D, \boldsymbol{x}_D\right)\right\rangle=\exp \left(-\textup{i} \varphi_i\right)\left\lvert\upsilon_i \left(t_S, \boldsymbol{x}_S\right)\right\rangle$, where $\varphi_i$ is the phase of oscillation. In flat spacetime, the phase is given by
\begin{equation}
 \varphi_i=E_i(t_D-t_s)-\boldsymbol{p}_i \cdot (\boldsymbol{x}_D-\boldsymbol{x}_s).  \label{eq4}
 \end{equation}
 The expression of phase $\varphi_i$ in curved spacetime, which is related to the properties of spacetime, can be represented in the covariant form as
\begin{equation}
\varphi_i=\int_S^Dp_\mu^{(i)}dx^\mu=\int_S^Dm_ig_{\mu\nu}\frac{dx^\nu}{ds}dx^\mu,\label{eq5}
\end{equation}
where $p_\mu^{(i)}$ is the canonical conjugate momentum to the coordinates $x^\mu$ and $ds$ denotes line element of the curved spacetime.

Neutrinos produced at the source $S(t_s,x_s)$ in a flavor eigenstate $\lvert\ \upsilon_\alpha\rangle$ travel to the mass eigenstate $\lvert\ \upsilon_i\rangle$, eventually recover the flavor eigenstate $\lvert\ v_\beta \rangle$ at the detector location $D$, thus neutrino flavor transition probability from initial produced $\alpha$ flavor to $\beta$ flavor at the detection point is given by \cite{retHSW}
\begin{equation}
\mathcal{P}_{\alpha\beta}\equiv\left\lvert\left\langle v_\beta\lvert v_\alpha(t_D,\boldsymbol{x}_D)\right\rangle\right\lvert^2=\sum_{i,j}U_{\beta i}U_{\beta j}^*U_{\alpha j}U_{\alpha i}^*exp[-\textup{i}(\varphi_i-\varphi_j)], \label{eq6}
\end{equation}
which means that when the initial and final flavor eigenstates are determined, the neutrino oscillation probability $\mathcal{P}_{\alpha\beta}$ depends only on the phase difference $\Delta\varphi_{ij}=\varphi_i-\varphi_j$, in the following section, we will focus on calculating the phase accumulation of neutrinos during their propagation in curved spacetime. Inspired by the work \cite{retBilenky:1978nj}, we assume that all the mass eigenstates in a flavour eigenstate initially produced at the source $S(t_s,x_s)$ have equal energy or momentum.

\section{Phase of neutrino oscillation in the EHB gravity around global monopole field}\label{sec4}

Now let's calculate relativistic neutrino oscillation phase described by the metric (Eq. \eqref{eq1}), for simplicity, the motion of neutrinos can be chosen to be confined on $\left(\theta=\frac{\pi}{2}\right)$ plane, thus the components of canonical momenta $p_\mu^{\left(h\right)}$ become
\begin{equation}
\begin{aligned}
 &p_t^{(h)}=m_hg_{tt}\frac{dt}{ds}=-m_hA\frac{dt}{ds}, \\  &p_r^{\left(h\right)}=m_hg_{rr}\frac{dr}{ds}=m_h\frac{1+a}{A}\frac{dr}{ds}, \\  &p_\phi^{\left(h\right)}=m_hg_{\phi\phi}\frac{d\phi}{ds}=m_hD\frac{d\phi}{ds},   \label{eq7}
\end{aligned}
\end{equation}
where $m_h$ is neutrino mass eigenstate, since the spacetime is static and vacuum, the energy and angular momentum of neutrinos are conserved so that $p_t^{(h)}=-E_h$ and $p_\phi^ {(h)}=J_h$, and the mass shell relation is given by
\begin{equation}
-m_{h}^2\equiv\ g^{tt}p_t^2+g^{rr}p_r^2+g^{\phi\phi}p_\phi^2, \label{eq8}
\end{equation}
in order to avoid confusion, in the following paragraphs we have removed the prefix $h$ in the momentum that indicates the mass eigenstate, and the corresponding calculation details can be found in the reference \cite{retChakrabarty}.

\subsection{radial propagating neutrino phase in a null trajectory $\left(d\phi=0\right)$}

We first examine the case of radial propagation of neutrinos, by substituting Eq. \eqref{eq7} and Eq. \eqref{eq8} into Eq. \eqref{eq5} reads
\begin{equation}
\varphi _ { h } = \int _ { r _ { S } } ^ { r _ { D } } \left[ - E _ { h } \left( \frac { d t } { d r } \right) _ { 0 } + \mathrm { p } _ { h } ( r ) \right] d r = \pm \int _ { r _ { S } } ^ { r _ { D } } \frac { E _ { h } } { A } \sqrt { 1 + a } \left[ - 1 + \sqrt { 1 - \frac { A m _ { h } ^ { 2 } } { E _ { h } ^ { 2 } } } \right] d r, \label{eq9}
\end{equation}
here the suffix 0 represents the null trajectory. Since only the gravitational field in the range of radial coordinates $r\in\left(2M,\infty\right)$ is considered, i.e. $0<A<1$, then we can expand the square root inside the bracket to get
\begin{equation}
\varphi _ { h } \approx \pm \int _ { r _ { s } } ^ { r _ { D } } \frac { E _ { h } } { A } \sqrt { 1 + a } \frac { A m _ { h } ^ { 2 } } { 2 E _ { h } ^ { 2 } } d r = \pm  \sqrt { 1 + a } \frac { m _ { h } ^ { 2 } } { 2 E _ { 0 }  } ( r _ { D } - r _ { s } ), \label{eq10}
\end{equation}
the above equation employs a relativistic approximation, which means that if $m_h\ll\ E_h$, we have $E _ { h } = \sqrt { P ^ { 2 } + m _ { h } ^ { 2 } } \approx E _ { 0 } + O \left( \frac { m _ { h } ^ { 2 } } { 2 E _ { 0 } } \right)$.
Observing the above equation, the background of ${\overline{\mu}}$ is not depend on the phase of the radial propagation of the neutrino, and if $a=0$, the phase will recover the case in the Schwarzschild spacetime \cite{retHSW}. Another interesting point is that Lorentz violating background affects the radial propagation of neutrinos even in the absence of a massive gravitational source $\left(M\rightarrow0\right)$, this is because Lorentz violating background also deforms spacetime.

\subsection{Non-radial propagating neutrino phase in a null trajectory($d\phi\neq0$)}

The neutrino phase travelling along a null trajectory (the details see \cite{retChakrabarty:2023kld}) in terms of the non-radial case can be written as
\begin{equation}
\begin{aligned}
\varphi_{h}=\int_{r_{S}}^{r_{D}} & {\left[-E_{h}\left(\frac{d t}{d r}\right)_{0}+\mathrm{p}_{h}(r)+J_{h}\left(\frac{d \phi}{d r}\right)_{0}\right] d r } \\
& = \pm \frac{m_{h}^{2}}{2 E_{0}} \int_{r_{S}}^{r_{D}} \sqrt{-g_{t t} g_{r r}}\left(1-\frac{b^{2}\left|g_{t t}\right|}{g_{\phi \phi}}\right)^{-\frac{1}{2}} d r \\
& = \pm \frac{m_{h}^{2}}{2 E_{0}} \sqrt{1+a} \int_{r_{S}}^{r_{D}} \sqrt{\frac{D}{D-b^{2} A}} d r
\end{aligned}, \label{eq11}
\end{equation}
where $b$ is the impact parameter. We can see that the phase is not only affected by the geometry of spacetime, but also by the trajectory of the neutrino from the propagation source to the detector for the case of $b\neq0$. Therefore, now we need to consider two cases discussing the phase:

(case 1) Neutrinos are produced at the source $S$, then propagate outwards non-radially without passing through dense celestial bodies until they reach the detector $D$. Since it is difficult to integrate Eq. \eqref{eq11} by replacing the metric coefficients, we rewrite it by using weak field approximation to get
\begin{equation}
 \varphi _ { h } \approx \pm \frac { m _ { h } ^ { 2 } } { 2 E _ { 0 } } \sqrt { 1 + a } \int _ { r _ { S } } ^ { r _ { D } } \left( \frac { r } { \sqrt { r ^ { 2 } - b ^ { 2 } ( 1 - \overline { \mu } ) } } - \frac { b ^ { 2 } M } { ( r ^ { 2 } - b ^ { 2 } ( 1 - \overline { \mu } ) ) ^ { \frac { 3 } { 2 } } } \right) d r
, \label{eq12}
\end{equation}
now the phase reads
\begin{equation}
 \varphi _ { h } ( r _ { s } \to r _ { D } ) = \frac { m _ { h } ^ { 2 } } { 2 E _ { 0 } } \sqrt { 1 + a } \left[ \sqrt { r _ { D } ^ { 2 } - b ^ { 2 } ( 1 - \overline { \mu } ) } - \sqrt { r _ { s } ^ { 2 } - b ^ { 2 } ( 1 - \overline { \mu } ) } \right. $$
$$ \left. + \frac { M } { ( 1 - \overline { \mu } ) } \left( \frac { r _ { D } } { \sqrt { r _ { D } ^ { 2 } - b ^ { 2 } ( 1 - \overline { \mu } ) } } - \frac { r _ { s } } { \sqrt { r _ { s } ^ { 2 } - b ^ { 2 } ( 1 - \overline { \mu } ) } } \right) \right] , \label{eq12}
\end{equation}
obviously, the above formula will reduce to that of the Schwarzschild case when $\overline{\mu}=a=0$ \cite{retNFO}.

(case 2) Neutrinos are produced at the source $S$, propagate towards the gravitational lens, and finally reaches the detector $D$, here $r=r_0$ represents the closest distance to the lens during the path. Meanwhile we apply the weak field approximation, thus the phase transforms into
\begin{equation}
\varphi _ { h } ( r _ { s } \to r _ { 0 } \to r _ { D } ) = \frac { m _ { h } ^ { 2 } } { 2 E _ { 0 } } \sqrt { 1 + a } \left[ \sqrt { r _ { D } ^ { 2 } - r _ { 0 } ^ { 2 } } + \sqrt { r _ { s } ^ { 2 } - r _ { 0 } ^ { 2 } } + \frac { M } { ( 1 - \overline { \mu } ) } \left( \sqrt { \frac { r _ { D } - r _ { 0 } } { r _ { D } + r _ { 0 } } } + \sqrt { \frac { r _ { s } - r _ { 0 } } { r _ { s } + r _ { 0 } } } \right) \right], \label{eq13}
\end{equation}
here $r_0$ can be determined by solving the equation
\begin{equation}
\left( \frac { d r } { d \phi } \right) _ { 0 } = \pm \frac { D } { b } \sqrt { \frac { 1 } { 1 + a } - \frac { b ^ { 2 } A } { D ( 1 + a ) } } = 0, \label{eq14}
\end{equation}
and its solution can be expressed as
\begin{equation}
r _ { 0 } \approx b \sqrt { 1 - \overline { \mu } } - \frac { M } { ( 1 - \overline { \mu } ) }. \label{eq15}
\end{equation}

After replacing $r_0$ with Eq.\eqref{eq15}, the Eq.\eqref{eq13}, when we consider the weak field approximation and $b\ \ll\ r_{s,D}$, becomes
\begin{equation}
\varphi _ { h } ( r _ { s } \to r _ { 0 } \to r _ { D } ) = \frac { m _ { h } ^ { 2 } } { 2 E _ { 0 } } \sqrt { 1 + a } ( r _ { D } + r _ { s } ) \left[ 1 + \frac { 2 M } { ( r _ { D } + r _ { s } ) ( 1 - \overline { \mu } ) } - \frac { b ^ { 2 } ( 1 - \overline { \mu } ) } { 2 r _ { D } r _ { s } } \right], \label{eq16}
\end{equation}
considering the Eq.\eqref{eq16}, we can see that whether the gravitational field exists or not, Lorentz symmetry breaking and global monopole field tend to cause the deformation of spacetime, thereby affecting the propagation phase, and obviously, and the phase in the case of weak gravitational lensing is the same with the one in Schwarzschild spacetime when $\overline{\mu}=a=0$ \cite{retHSW}.

\section{Neutrino oscillation probability in the EHB gravity around global monopole field}\label{sec5}

\begin{figure}[b]
\centerline{\includegraphics[width=12.0cm]{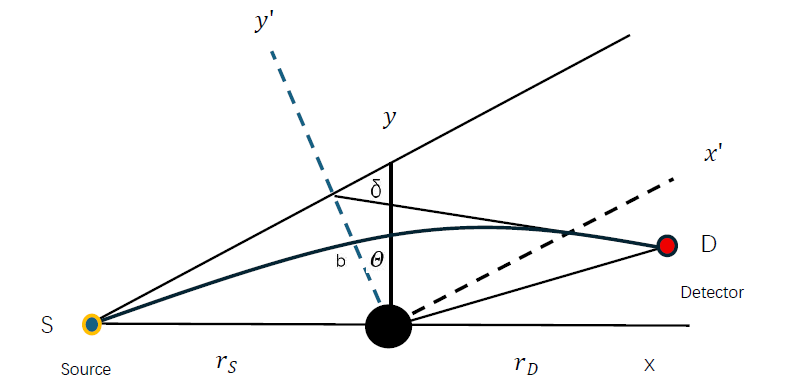}}
\caption{Diagrammatic representation of weak lensing of neutrinos.}
\label{fig1}
\end{figure}

As depicted in Fig. \ref{fig1}, neutrinos are produced in a same source and propagate outward along different paths. Due to the presence of a strong gravitational field (Schwarzschild-like black hole), the trajectories of some neutrinos deviate and converge at a common detection point $D$, resulting in the interference phenomenon, which is similar to how a convex lens (equivalent to a black hole) can converge the optical rays of the sun (equivalent to the propagation of neutrinos). Considering the fact that neutrinos traveling through different paths $p$ from S to D, so the neutrino flavour eigenstate need to be expressed as \cite{retChakrabarty}
\begin{equation}
|v_{\alpha}(t_{D}, \boldsymbol{x}_{D})\rangle = N \sum_{i} U_{\alpha i}^{*} \sum_{p} \exp(-\textup{i} \varphi_{i}^{p}) |v_{i}(t_{s}, \boldsymbol{x}_{s})\rangle, \label{eq17}
\end{equation}
where $\varphi_i^p$ is given by Eq.\eqref{eq16} and $N$ is a normalization factor. Considering the transition from an initial flavor eigenstate $\alpha$ at the source $S$ to a final  flavor eigenstate $\beta$ at the detector $D$, the oscillation probability can be expressed as
\begin{equation}
 \mathcal { P } _ { \alpha \beta } ^ { l e n s } \equiv \left| \left\langle v _ { \beta } | v _ { \alpha } \left( t _ { D }, \boldsymbol { x } _ { D } \right) \right\rangle \right| ^ { 2 } = | N | ^ { 2 } \sum _ { i, j } U _ { \beta i } U _ { \beta j } ^ { * } U _ { \alpha j } U _ { \alpha i } ^ { * } \sum _ { p, q } \exp \left( -\textup{i} \Delta \varphi _ { i j } ^ { p q } \right), \label{eq18}
\end{equation}
the normalization constant $N$ is expressed as
\begin{equation}
|N|^{2}=\left(\sum_{i}\left|U_{\alpha i}\right|^{2} \sum_{p, q} \exp \left( -\textup{i} \Delta \varphi_{i i}^{p q}\right)\right)^{-1}. \label{eq19}
\end{equation}

As mentioned before, when the initial and final flavor eigenstates are determined, the oscillation probability $\mathcal { P } _ { \alpha \beta } ^ { l e n s }$ depends on the phase difference, thus in order to facilitate analysis, $\Delta\varphi_{ij}^{pq}$ can be represented by Eq.\eqref{eq16} as
\begin{equation}
{\Delta\varphi}_{ij}^{pq}=\varphi_i^p-\varphi_j^q=\Delta m_{ij}^2A_{pq}+\Delta b_{pq}^2B_{ij}, \label{eq20} \end{equation}
where
\begin{equation}
\begin{aligned}
 &A_{p q}=\frac{\sqrt{1+a}}{2 E_{0}}\left(r_{D}+r_{s}\right)\left[1+\frac{2 M}{\left(r_{D}+r_{s}\right)(1-\bar{\mu})}-\frac{(1-\bar{\mu}) \sum b_{p q}^{2}}{4 r_{D} r_{s}}\right], \\  & B_{i j}=-\frac{(1-\bar{\mu}) \sqrt{1+a}}{8 E_{0}} \sum m_{i j}^{2}\left(\frac{1}{r_{D}}+\frac{1}{r_{S}}\right),  \\  & \sum m _ { i j } ^ { 2 } = m _ { i } ^ { 2 } + m _ { j } ^ { 2 }, \\& \Delta b_{pq}^2=b_p^2-b_q^2,  \\&  \sum b_{pq}^2=b_p^2+b_q^2, \\ &\Delta m _ { i j } ^ { 2 } = m _ { i } ^ { 2 } - m _ { j } ^ { 2 },
\label{eq21}
\end{aligned}
\end{equation}
Observing this form, we can see that if $p\neq q$, it is possible for the oscillation probability $\mathcal { P } _ { \alpha \beta } ^ { l e n s }$ including the terms $\sum m_{ij}^2$ and $\Delta m _ { i j } ^ { 2 }$, therefore, in principle, the observation of neutrinos in the lens could potentially detect the information about their absolute mass \cite{retChakrabarty}.

Now our analysis focus on examining the effects of gravitational lensing on the neutrino flavor transformation aiming to analyze the roles of the parameters $\overline{\mu}$  and $a$, in order to understand the oscillation probability qualitatively and quantitatively in the EHB gravity around global monopole field, we take into account a transition from $\upsilon_e$ to $\upsilon_\mu$, thus the oscillation probability is given by
\begin{equation}
\begin{aligned}
\mathcal{P} _ { e \mu } ^ { l e n s } &= | N | ^ { 2 } \sin ^ { 2 } 2 \gamma \bigg\{
    \frac{1}{2} \cos\bigl( \Delta b_{12}^2 B_{11} \bigr) + \frac{1}{2} \cos\bigl( \Delta b_{12}^2 B_{22} \bigr) + \sin^2\bigl( \frac{\Delta m_{21}^2 A_{11}}{2} \bigr) \\
    &\quad + \sin^2\bigl( \frac{\Delta m_{21}^2 A_{22}}{2} \bigr) - \cos\bigl( \Delta m_{21}^2 A_{12} \bigr) \cos\bigl( \Delta b_{12}^2 B_{12} \bigr)
\bigg\},
\label{eq22}
\end{aligned}
\end{equation}
where $|N|^{2}=\frac{1}{2}\left[1+\cos ^{2} \gamma \cos \left(\Delta b_{12}^{2} B_{11}\right)+\sin ^{2} \gamma \cos \left(\Delta b_{12}^{2} B_{22}\right)\right]^{-1}$ and $\gamma$ is the mixing angle \cite{retEsteban:2018azc}, we can see that $\mathcal{P}_{\alpha\beta}^{lens}$ depends on the neutrino mass ordering and for $\Delta m^2>0$ and $\Delta m^2<0$, it results in different results ($\gamma\neq\frac{\pi}{4}$).

\section{Numerical results: two flavor case}\label{sec6}

We prefer to solve the path-dependent impact parameter $b_p$ in terms of the geometric quantities of the system. As shown in Fig. \ref{fig1}, it shows the description of lensing phenomena, in which the physical distance from the source to the lens is $r_S$ and the physical distance from the lens to the detector is $r_D$. For convenience, a new coordinate system $(x^\prime,y^\prime)$ is introduced by rotating the $(x, y)$ coordinate system  through an angle $\Theta$ such that $x^\prime=x\cos{\Theta}+y\sin{\Theta}$ and $y^\prime=y\cos{\Theta}-x\sin{\Theta}$.
Inspired by the work \cite{retCASA}, the angle of deflection $\delta$ of neutrinos  in the $(x^\prime, y^\prime)$ coordinate system can be expressed as
\begin{equation}
 \delta \sim \frac { y _ { D } ^ { \prime } - b } { x _ { D } ^ { \prime } } = - \frac { 4 M } { b ( 1 - \overline { \mu } ) } - \frac { \overline { \mu } \pi } { 2 } - \frac { a \pi } { 2 }, \label{eq23}
\end{equation}
where $\left(x_D^\prime,y_D^\prime\right)$ denotes the location of the detector and it should be noted that this equation we take consideration of the weak lensing limit. Now by using the identity $ \sin{\Theta}=\frac{b}{r_s}$ from Fig. \ref{fig1}, Eq.\eqref{eq23} can be reformulated as
\begin{equation}
b^{2}\left( \frac { x _ { D } } { r _ { s } } + 1 - \frac { y _ { D } \pi } { 2 r _ { s } } ( a + \overline { \mu } ) \right) - \frac { 4 M y _ { D } b } { ( 1 - \overline { \mu } ) r _ { s } } = \sqrt { 1 - \left( \frac { b } { r _ { s } } \right) ^ { 2 } } \left( y _ { D } b + ( a + \overline { \mu } ) \frac { x _ { D } \pi b } { 2 } + \frac { 4 M } { ( 1 - \overline { \mu } ) } x _ { D } \right). \label{eq24}
\end{equation}

\begin{figure}[htbp]
\centerline{\includegraphics[width=17.0cm]{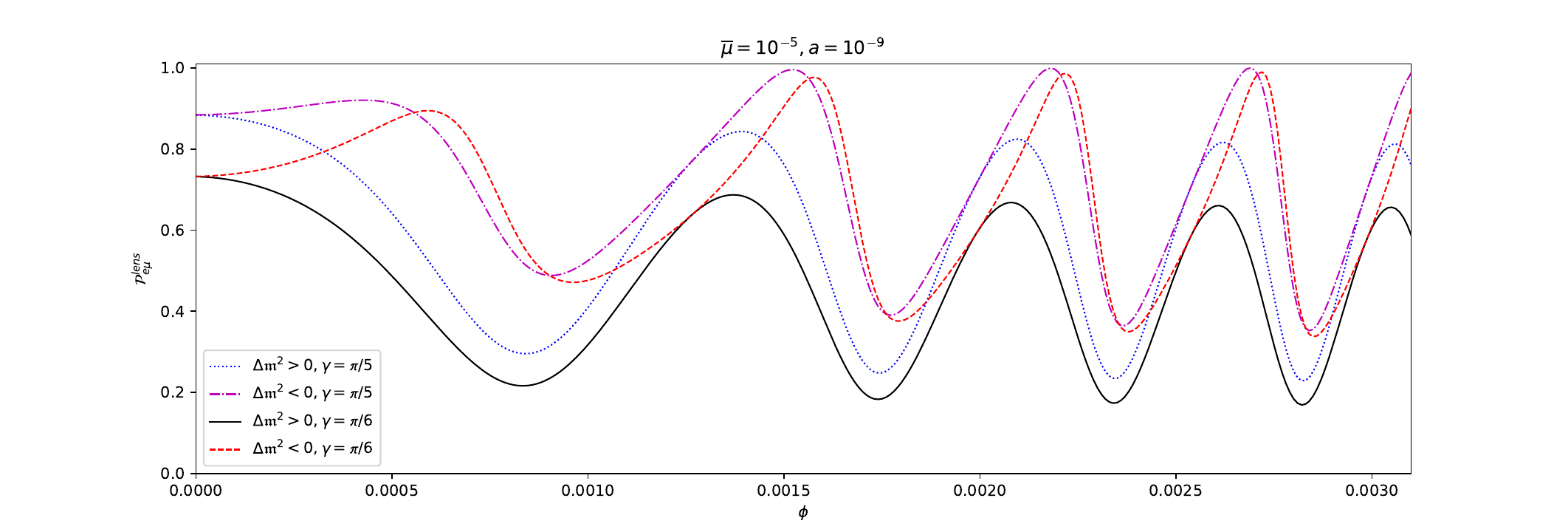}}
\caption{Oscillation probability of gravitationally lensed neutrinos. Values of the parameters are as follows: $M=M_{\odot}$, $\Delta m^2={10}^{-3}{eV}^2$ and the mass of the lightest neutrino is zero.}
\label{fig2}
\end{figure}

\begin{figure}[htbp]
\begin{center}
\begin{tabular}{cc}
\begin{minipage}[t]{0.45\linewidth}
\centerline{\includegraphics[width=17.0cm]{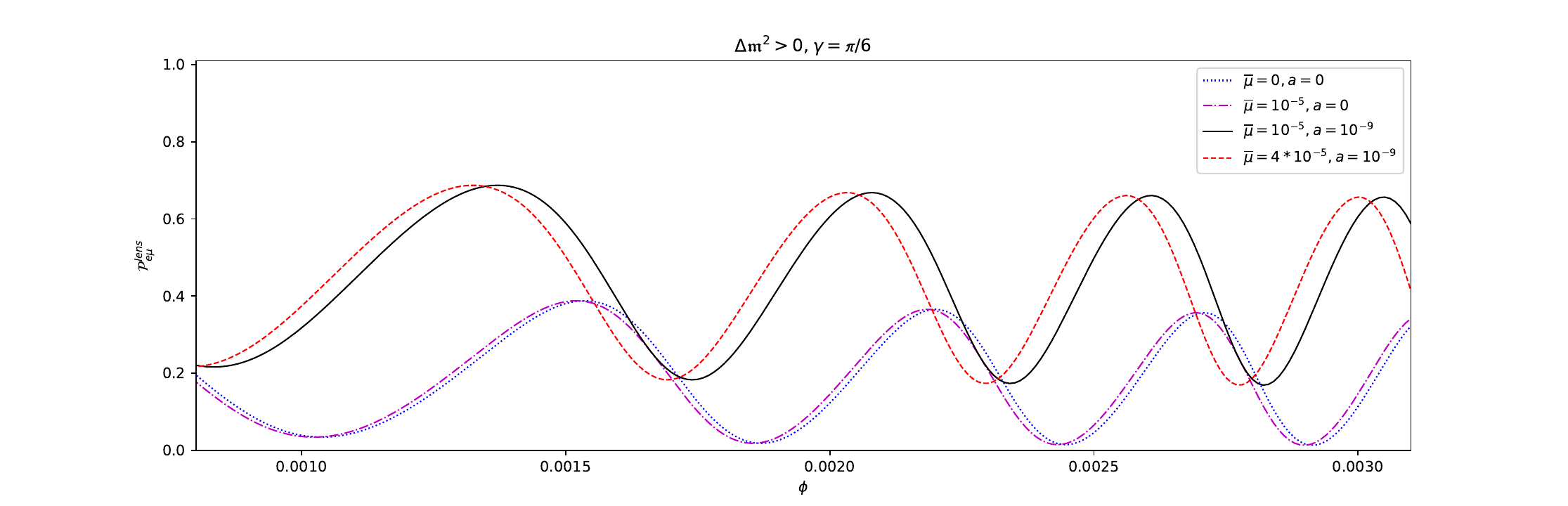}}
\end{minipage}
\\
\begin{minipage}[t]{0.45\linewidth}
\centerline{\includegraphics[width=17.0cm]{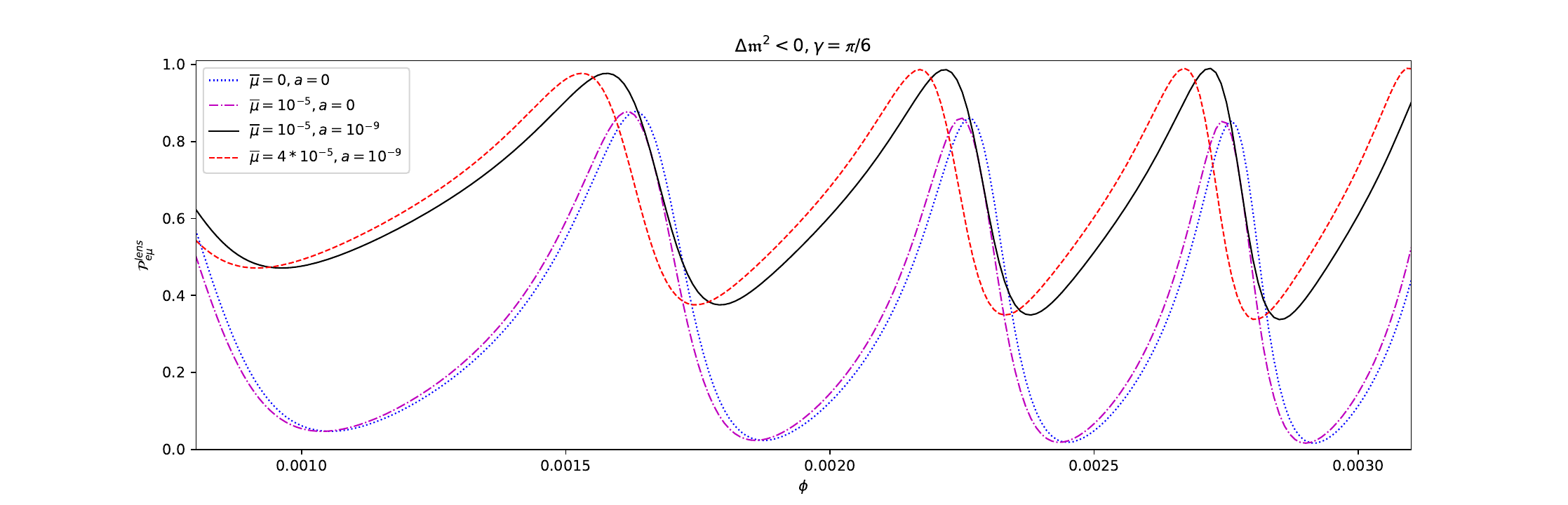}}
\end{minipage}
\end{tabular}
\caption{Oscillation probability of gravitationally lensed neutrinos. Values of the parameters are as follows:$M=M_{\odot}$, $\Delta m^2={10}^{-3}{eV}^2$ and the mass of the lightest neutrino is zero.}
\label{fig3}
\end{center}
\end{figure}

\begin{figure}[htbp]
\begin{center}
\begin{tabular}{cc}
\begin{minipage}[t]{0.45\linewidth}
\centerline{\includegraphics[width=17.0cm]{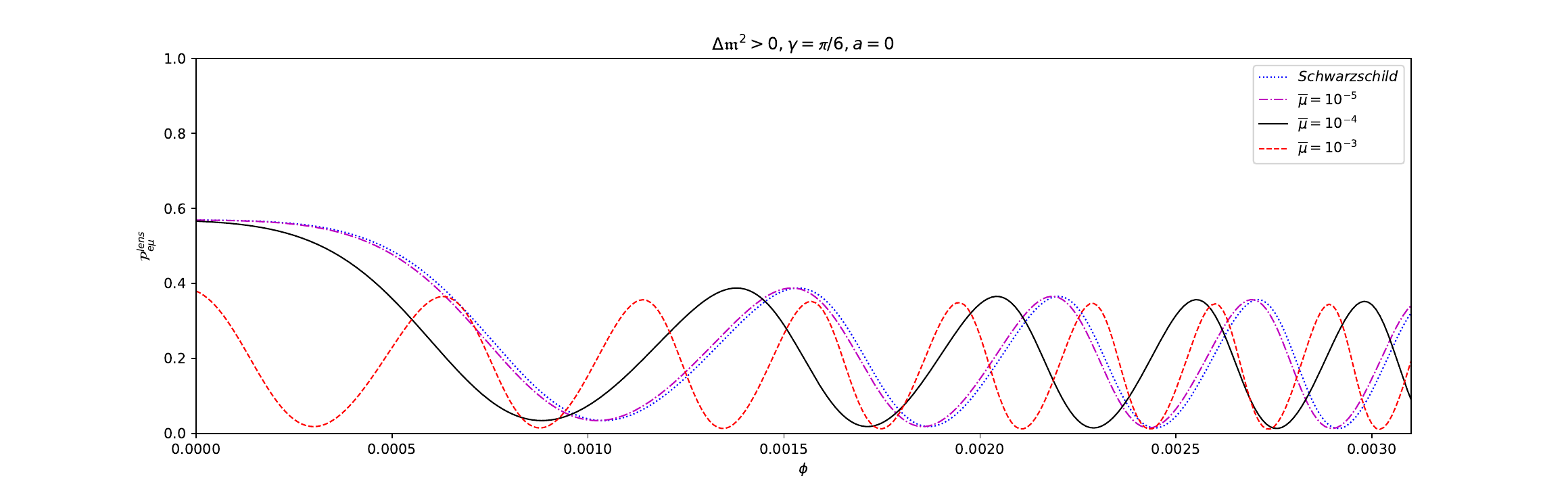}}
\end{minipage}
\\
\begin{minipage}[t]{0.45\linewidth}
\centerline{\includegraphics[width=17.0cm]{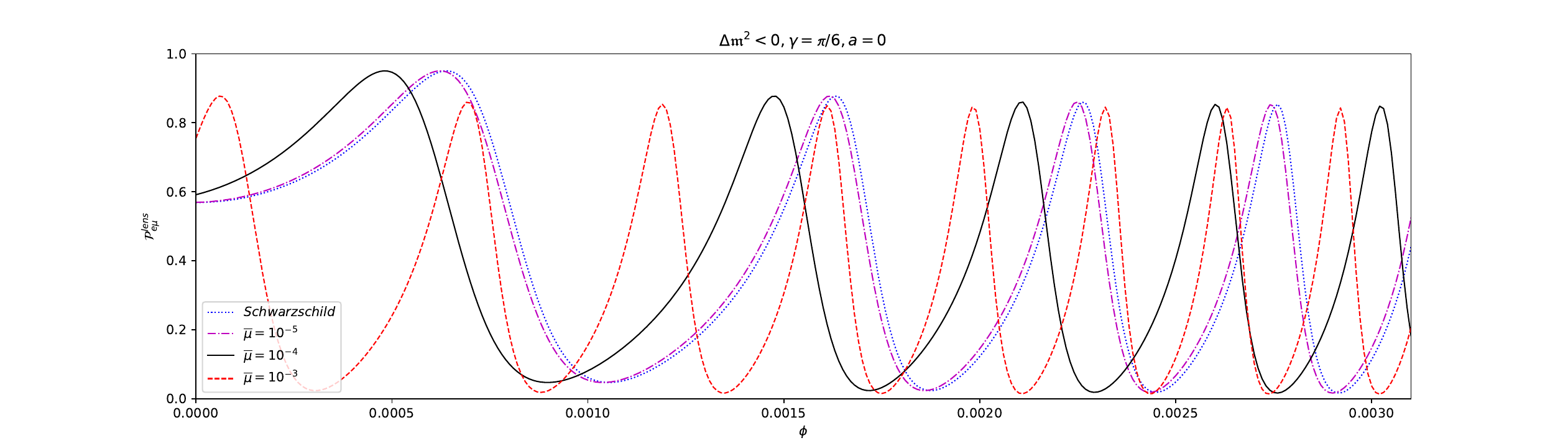}}
\centerline{(a)}
\end{minipage}
\\
\begin{minipage}[t]{0.45\linewidth}
\centerline{\includegraphics[width=17.0cm]{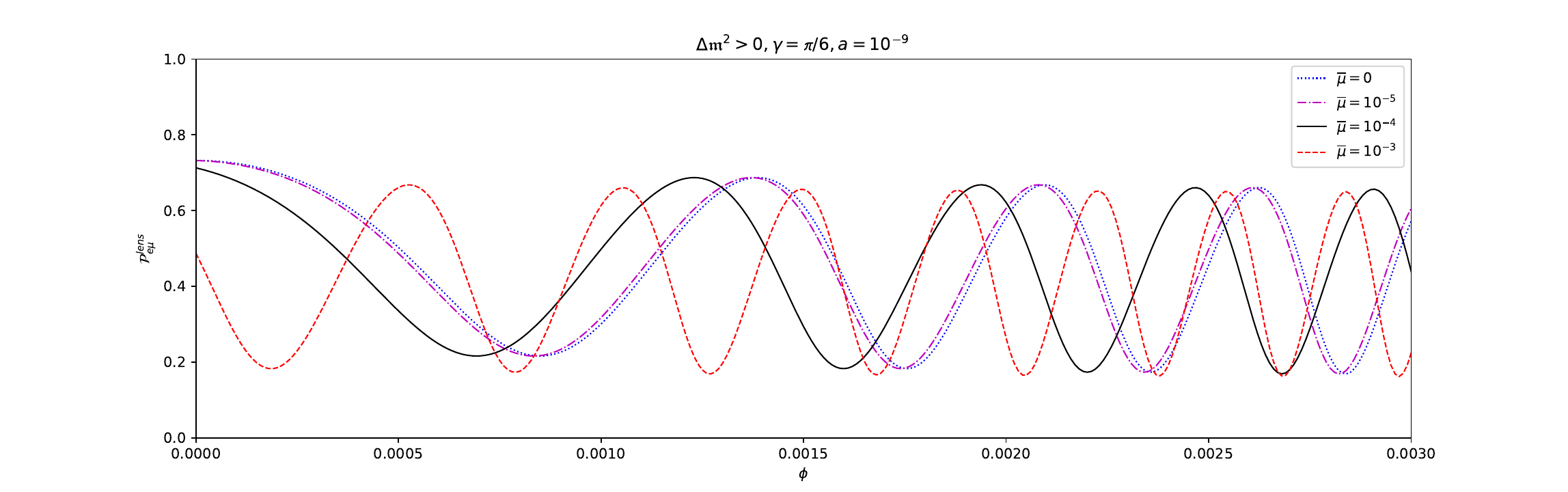}}
\end{minipage}
\\
\begin{minipage}[t]{0.45\linewidth}
\centerline{\includegraphics[width=17.0cm]{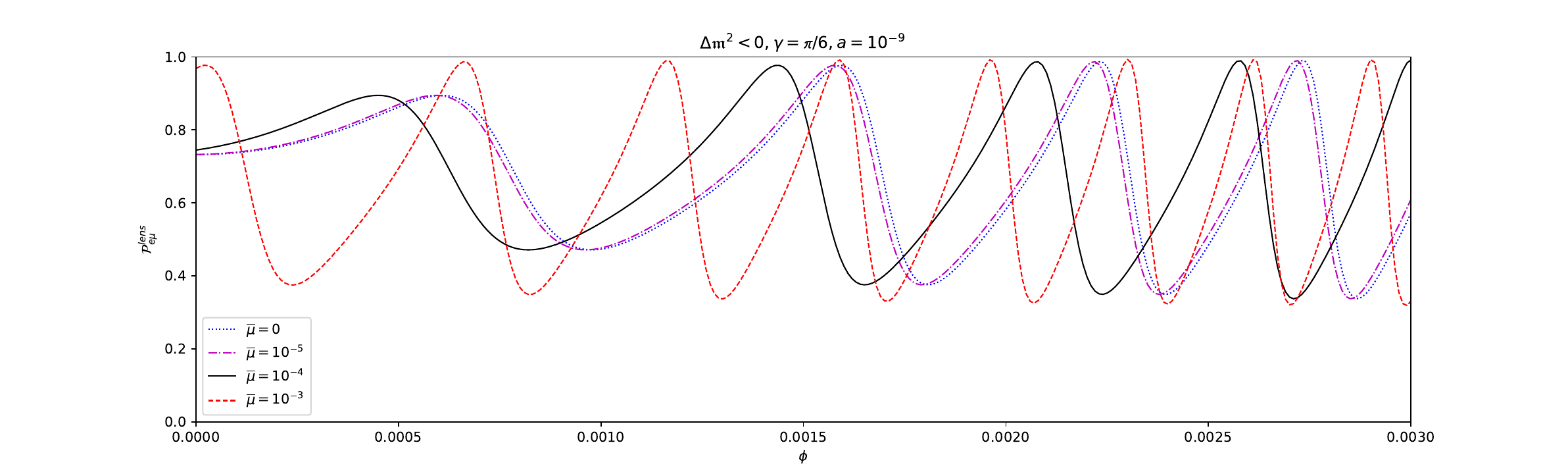}}
\centerline{(b)}
\end{minipage}
\end{tabular}
\caption{Effect of the global monopole $\overline { \mu }$ on the oscillation probability. Values of the parameters are as follows:$M=M_{\odot}$, $\Delta m^2={10}^{-3}{eV}^2$ and the mass of the lightest neutrino is zero.}
\label{fig4}
\end{center}
\end{figure}

\begin{figure}[htbp]
\begin{center}
\begin{tabular}{cc}
\begin{minipage}[t]{0.45\linewidth}
\centerline{\includegraphics[width=17.0cm]{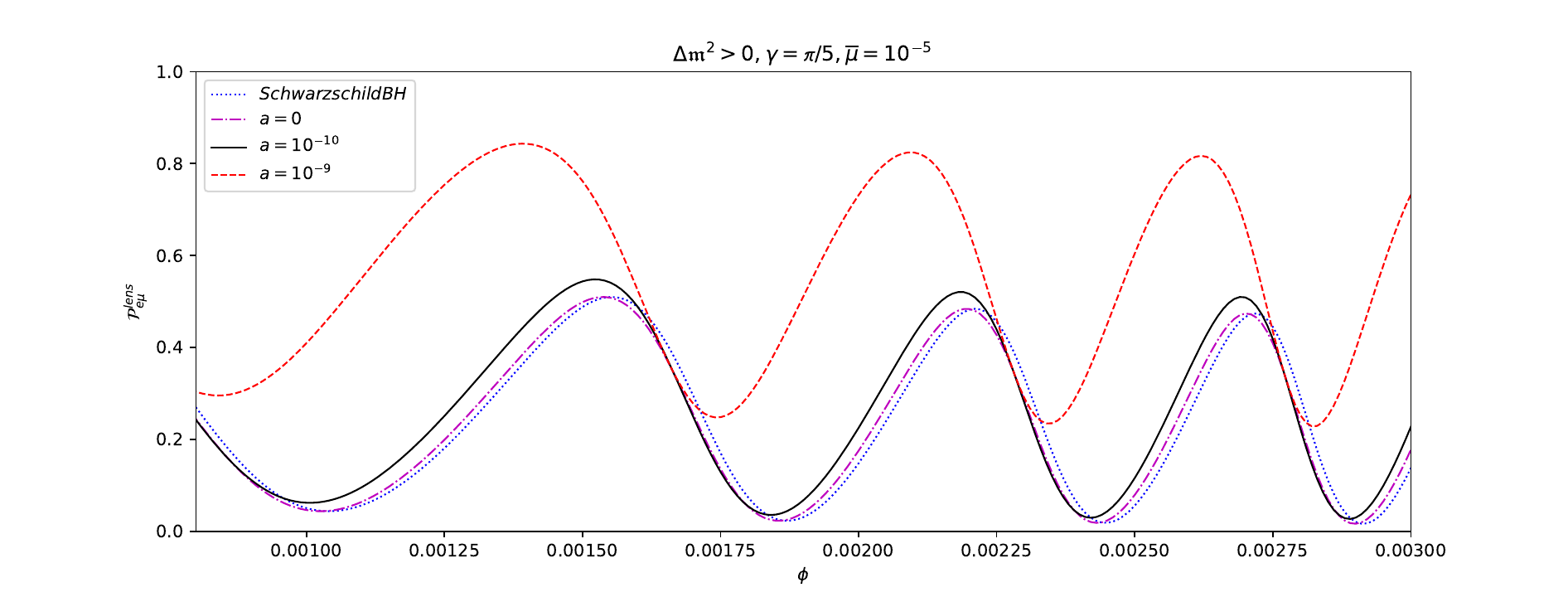}}
\end{minipage}
\\
\begin{minipage}[t]{0.45\linewidth}
\centerline{\includegraphics[width=17.0cm]{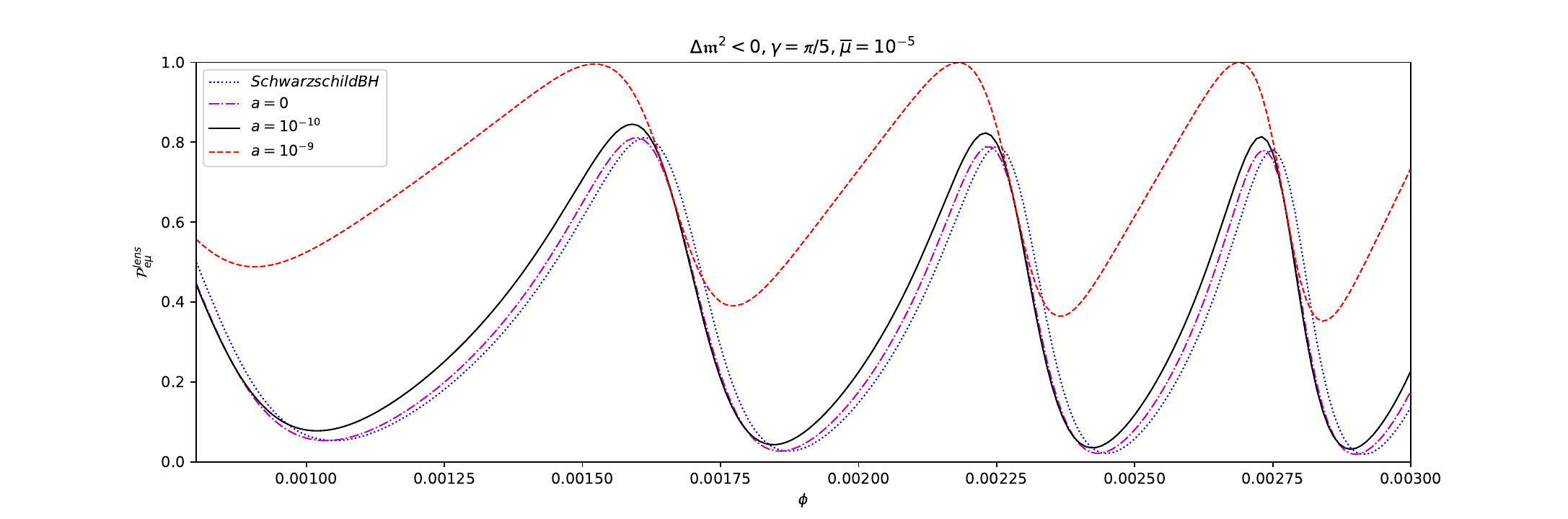}}
\end{minipage}
\end{tabular}
\caption{Effect of the Lorentz-violating parameter $a$ on the oscillation probability. Values of the parameters are as follows:$M=M_{\odot}$, $\Delta m^2={10}^{-3}{eV}^2$ and the mass of the lightest neutrino is zero.}
\label{fig5}
\end{center}
\end{figure}

\begin{figure}[htbp]
\begin{center}
\begin{tabular}{cc}
\begin{minipage}[t]{0.45\linewidth}
\centerline{\includegraphics[width=17.0cm]{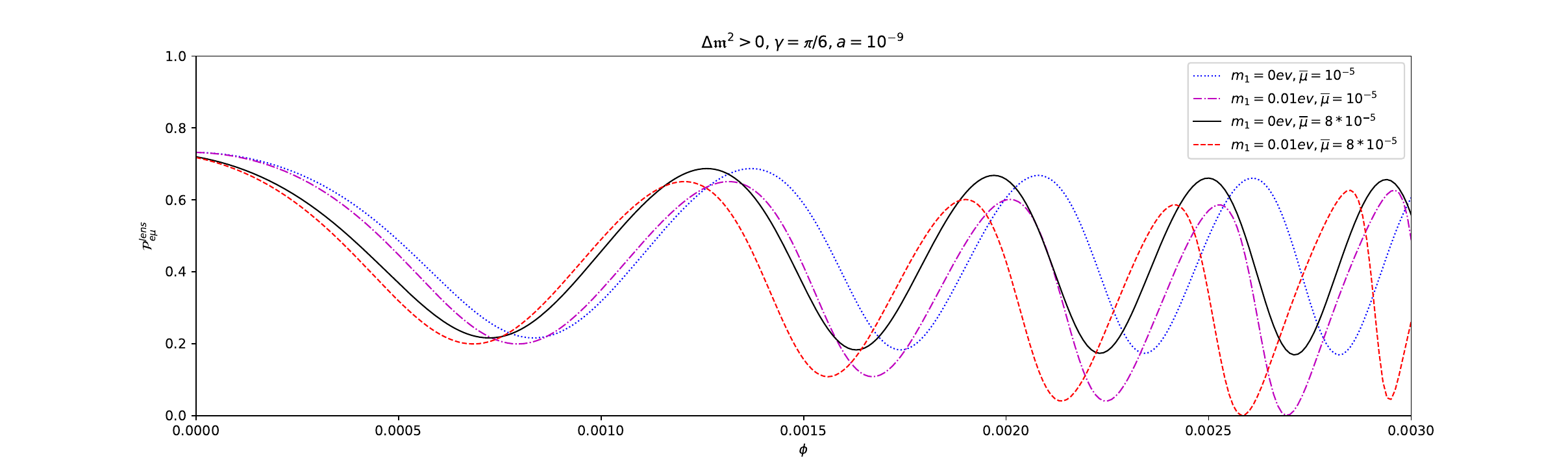}}
\centerline{(a)}
\end{minipage}
\\
\begin{minipage}[t]{0.45\linewidth}
\centerline{\includegraphics[width=17.0cm]{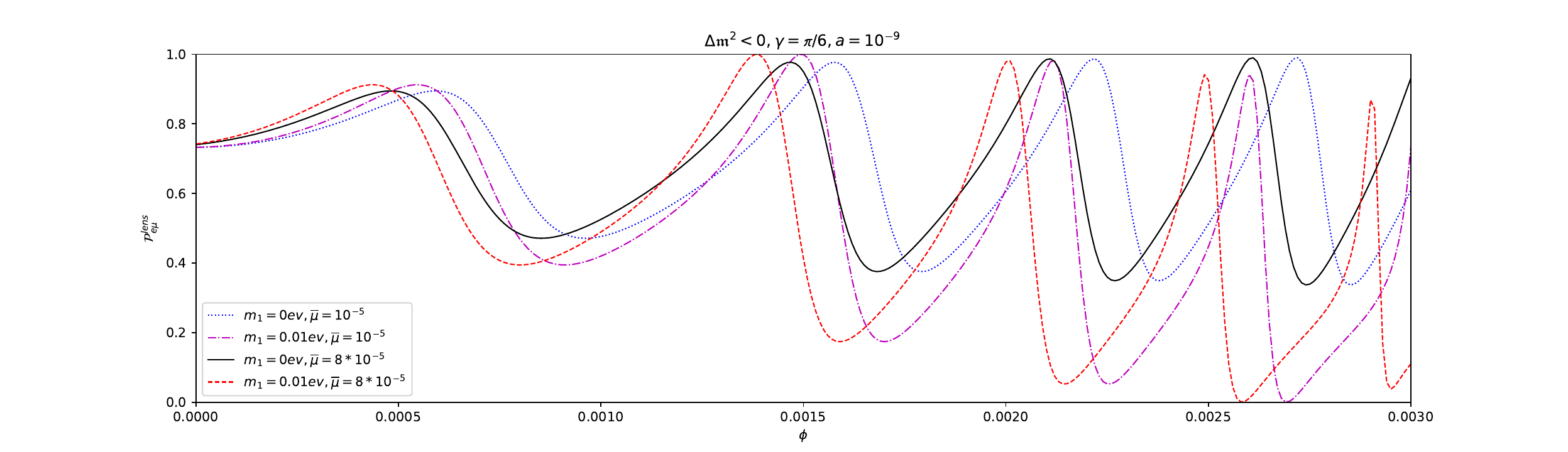}}
\centerline{(b)}
\end{minipage}
\end{tabular}
\caption{Oscillation of probability of the lightest neutrino without mass and with mass. Values of the parameters are as follows:$M=M_{\odot}$ and $\Delta m^2={10}^{-3}{eV}^2$. }
\label{fig6}
\end{center}
\end{figure}

Next we shall calculate the oscillation probability of neutrinos that are lensed by a Schwarzschild-like black hole in the EHB gravity around global monopole field. In order to have a quantitative understanding of oscillation probability, we consider a Sun-Earth system with the Sun as the lens and the Earth as a detector $(r_D=\ {10}^8km)$. Assuming that the gravitational field of the Sun is represented by the exterior of the Schwarzschild-like metric in the EHB gravity around global monopole field in the weak field limit, the source is at a distance $(r_s={10}^5r_D)$ from the sun and it emits high-energy neutrinos with $E_0=\ 10MeV$. Then we assume that the detector follows a circular orbit around the sun ($x_D\ =\ r_D\cos{\phi}$, $y_D=r_D\sin{\phi}$), thus by dealing with Eq.\eqref{eq24} numerically to obtain the impact parameters, we can derive the oscillation probability of neutrinos with respect to the azimuth angle $\phi$. The oscillation probability for the two flavor toy model of neutrinos is shown in Fig. \ref{fig2}, \ref{fig3}, \ref{fig4}, \ref{fig5} and \ref{fig6} as a function of the azimuth angle $\phi$, our main purpose is the dependence of oscillation probability on global monopole $\overline { \mu }$.

Fig. \ref{fig2} shows the dependence of oscillation probability on the neutrino mass ordering for the nonmaximal $\gamma$, which indicates that for the nonmaximal $\gamma$, the probability in the case of inverted hierarchy is large than the normal hierarchy and when the nonmaximal $\gamma$ transitions from $\gamma=\frac{\pi}{6}$ to $\gamma=\frac{\pi}{5}$, the probability of the normal hierarchy increases. It should be noted that the maximal mixing angle is set to be $\gamma=\frac{\pi}{4}$, interestingly, the probability in the case of inverted hierarchy is almost the same as that in the normal hierarchy if $\gamma=\frac{\pi}{4}$, and when $\gamma>\frac{\pi}{4}$, the probability in the case of inverted hierarchy is lower than the normal hierarchy, actually, these behaviors can be verified in Eq.\eqref{eq22}. On the other hand, inpired by the works \cite{retCASA,retVilenkin,retPantig:2024kqy}, the selection of the Lorentz-violating parameter $a$ and the global monopole $\overline { \mu }$ are set according to the constraints in some key classical tests, such a selection is to better fit the actual situation of the experiment.

Fig. \ref{fig3} shows the deviation of the oscillation probability of neutrinos in Schwarzschild-like spacetime in the EHB gravity around global monopole field from the case in Schwarzschild black hole, which indicates that when the Lorentz-violating parameter $a$ is absent, the oscillation probability has a significant deviation from the Schwarzschild case for sufficiently large values of the global monopole $\overline { \mu }$, and while when we consider the Lorentz-violating parameter $a$ and global monopole $\overline { \mu }$ simultaneously, this deviation becomes even more obvious. From the oscillation probability given by Eq.\eqref{eq22} and Fig. \ref{fig3}, we can see that the Lorentz-violating parameter $a$ and the global monopole $\overline { \mu }$ indeed affect the oscillation probability, however, how the two parameters affect the oscillation probability? We depict Fig. \ref{fig4} and Fig. \ref{fig5} to explain this problem.

Fig. \ref{fig4} (a) shows the effect of the global monopole $\overline { \mu }$ on the oscillation probability in the absence of the Lorentz-violating parameter $a$, which indicates that     the global monopole field has a very small effect on the peak of the oscillation probability, it mainly plays a role in changing the frequency of the oscillations of probabilities, i.e. as the global monopole $\overline { \mu }$ increases, the frequency of the oscillation of probability increases, and as depicted in Fig. \ref{fig4} (b), this conclusion remains valid even if the Lorentz-violating parameter $a$ exists.

Fig. \ref{fig5} shows the effect of the Lorentz-violating parameter $a$ on the oscillation probability in the presence of the global monopole $\overline { \mu }$, which indicates that the increasing Lorentz-violating parameter $a$ increases the oscillation probability, and if the global monopole $\overline { \mu }$ is absent, the same result has been verified in \cite{retShi:2025plr}.

On the other hand, according to the work \cite{retHSW}, the frequency of the oscillations of probabilities increases as the lightest neutrino mass, which implies that the lightest neutrino mass also plays a role in regulating the frequency of the oscillations of probabilities. In this line of thought, we compare the oscillation probability of the lightest neutrino with mass and without mass in the EHB gravity around global monopole field in Fig. \ref{fig6}, it shows that when $\overline { \mu }=10^{-5}$ or $\overline { \mu }=8\times 10^{-5}$, as the lightest neutrino mass increases from 0 to 0.01 eV, the frequency of the oscillations of probabilities will increase, especially when the lightest neutrino mass increases to 0.01 eV, the frequency of the oscillations of probabilities still increases with the increase of the global monopole $\overline { \mu }$ (see the red dashed line and the purple dotted-dashed line), which means that the lightest neutrino mass and the global monopole $\overline { \mu }$ have the same mechanism influencing the frequency of the oscillations of probabilities, however this also makes it difficult for us to use the neutrino oscillation phenomenon as a messenger to detect global monopole.

\section{Conclusion}\label{sec7}

This work we studied the neutrino behavior in the EHB gravity around global monopole field and discussed how the Lorentz-violating parameter $a$ and the global monopole $\overline { \mu }$ affect neutrino behavior in this curved spacetime.  Firstly, we derived the expression of the neutrino oscillation phase in the EHB gravity around global monopole field in weak field limit, the oscillation probabilities for neutrinos lensed by the gravitational source are obtained, finally, by plotting the probability of $\upsilon_e \rightarrow \upsilon_\mu$ as a function of the azimuth angle $\phi$ for the normal and inverted ordering of neutrino masses, we qualitatively discussed the effect of the geometry of the spacetime of the EBH black holes on the neutrino oscillation probability and compared the neutrino oscillation probability with the case of lensing by Schwarzschild black hole. Our results show that due to the effect of the Lorentz-violating parameter $a$ and the global monopole $\overline { \mu }$, the curve of the oscillation probability in the EHB gravity around global monopole field and the curve of the oscillation probability in the classic Schwarzschild black hole do not overlap, thus we may view that it offers us a new option by using the gravitational effect of neutrinos to detect celestial bodies, meanwhile, since the neutrino oscillation probability in the curved spacetime contains the terms $\sum m_{ij}^2$ and $\Delta m_{ij}^{2}$,  it is also possible to use the known spacetime geometry estimating the absolute mass of a single neutrino.

Another important point is that the effects of the Lorentz-violating parameter $a$ and the global monopole $\overline { \mu }$ on oscillation probability are different. Specifically,  as the Lorentz-violating parameter $a$ increases, the peak of the oscillation probability will increase, this obvious difference is expected to meet the current observation accuracy of Icecube-Gen2. Besides, the increase of the global monopole $\overline { \mu }$ and the lightest neutrino mass will both result in an increase for the frequency of the oscillations of probabilities, the effects of them on the oscillation probability are the same, thus it is a big challege for us to use the neutrino oscillation phenomenon as a messenger detecting the global monopole. Fortunately, we have other established techniques for observing physical celestial bodies, including the gravitational wave \cite{retWu:2022eiv}, black hole shadow \cite{retWu:2024hxr} and so on, they can serve as technical supplements, making the measurement of the global monopole $\overline { \mu }$ possible and improving the detection mechanism for compact celestial bodies.

\section*{Acknowledgments}
This research was funded by the Guizhou Provincial Science and Technology Project(Guizhou Scientific Foundation-ZK[2022] General 491), the doctoral startup fund of Guiyang University(GYU-KY-(2025)-06) and the National Natural Science Foundation of China (Grant No.12265007) .





\begin{thebibliography}{77}

\bibitem{retMLO} M.~Losada, Y.~Nir, G.~Perez, I.~Savoray and Y.~Shpilman, Parametric resonance in neutrino oscillations induced by ultra-light dark matter and implications for KamLAND and JUNO. JHEP \textbf{03}, 032 (2023)
\bibitem {retYFU} Y.~Fukuda \textit{et al.}, Evidence for oscillation of atmospheric neutrinos. Phys. Rev. Lett. \textbf{81}, 1562 (1998)

\bibitem{retMTA} M.~Tanabashi \textit{et al.} [Particle Data Group], Review of Particle Physics. Phys. Rev. D \textbf{98}, 030001 (2018)

\bibitem{retCYC} C.~Y.~Cardall and G.~M.~Fuller, Neutrino oscillations in curved space-time: An Heuristic treatment.
Phys. Rev. D \textbf{55}, 7960  (1997)

\bibitem{retNFO} N.~Fornengo, C.~Giunti, C.~W.~Kim and J.~Song, Gravitational effects on the neutrino oscillation.
Phys. Rev. D \textbf{56}, 1895 (1997)

\bibitem{retJGP} J.~G.~Pereira and C.~M.~Zhang, Some remarks on the neutrino oscillation phase in a gravitational field.
Gen. Rel. Grav. \textbf{32}, 1633 (2000)

\bibitem{retGLA} G.~Lambiase, G.~Papini, R.~Punzi and G.~Scarpetta, Neutrino optics and oscillations in gravitational fields.
Phys. Rev. D \textbf{71}, 073011 (2005)


\bibitem{retSIG} S.~I.~Godunov and G.~S.~Pastukhov, Neutrino Oscillations in Gravitational Field.
Phys. Atom. Nucl. \textbf{74}, 302 (2011)

\bibitem{retSCH} S.~Chakraborty, Aspects of Neutrino Oscillation in Alternative Gravity Theories.
JCAP \textbf{10}, 019 (2011)

\bibitem{retLBU} L.~Buoninfante, G.~G.~Luciano, L.~Petruzziello and L.~Smaldone, Neutrino oscillations in extended theories of gravity.
Phys. Rev. D \textbf{101}, 024016 (2020)

\bibitem{retKBO} K.~Boshkayev, O.~Luongo and M.~Muccino, Neutrino oscillation in the $q$-metric.
Eur. Phys. J. C \textbf{80}, 964 (2020)


\bibitem{retACA} A.~Capolupo, G.~Lambiase and A.~Quaranta, Neutrinos in curved spacetime: Particle mixing and flavor oscillations.
Phys. Rev. D \textbf{101}, 095022 (2020)
\bibitem{retHSW} H.~Swami, K.~Lochan and K.~M.~Patel, Signature of neutrino mass hierarchy in gravitational lensing.
Phys. Rev. D \textbf{102}, 024043 (2020)

\bibitem{retChakrabarty} H.~Chakrabarty, D.~Borah, A.~Abdujabbarov, D.~Malafarina and B.~Ahmedov, Effects of gravitational lensing on neutrino oscillation in $ \gamma $-spacetime. Eur. Phys. J. C \textbf{82}, 24 (2022)

\bibitem{retJacobson} T.~Jacobson and D.~Mattingly, Gravity with a dynamical preferred frame. Phys. Rev. D \textbf{64}, 024028 (2001)

\bibitem{retHorava} P.~Horava, Quantum Gravity at a Lifshitz Point. Phys. Rev. D \textbf{79}, 084008 (2009)

\bibitem{retCASA} Casana, R., Cavalcante, A., Poulis, F. P. and Santos, E. B., Exact Schwarzschild-like solution in a bumblebee gravity model. Phys. Rev. D \textbf{97}, 104001 (2018)



\bibitem{retSenjaya:2025cgk} D.~Senjaya, Scalar quasibound states in Einstein-Maxwell-Bumblebee black holes with non-minimal Maxwell-Bumblebee coupling. Nucl. Phys. B \textbf{1022}, 117226 (2026)

\bibitem{retSingh:2025fuv} Y.~P.~Singh, J.~Choudhury, T.~I.~Singh and D.~J.~Gogoi, Field perturbations and observables in a charged black hole within bumblebee gravity. Eur. Phys. J. Plus \textbf{140}, 1118  (2025)

\bibitem{retYuChih:2025hsg} H.~Y.~YuChih and Y.~Shen, Energy extraction from a rotating black hole via magnetic reconnection: Bumblebee gravity. Phys. Rev. D \textbf{112}, 104016 (2025)

\bibitem{retSekhmani:2025gvv} Y.~Sekhmani, S.~K.~Maurya, J.~Rayimbaev, M.~Altanji, I.~Ibragimov and S.~Muminov, Exact static and slowly rotating phantom conformal nonlinear-electrodynamic black holes in Bumblebee gravity with Lorentz Violation and shadow phenomenology. Phys. Dark Univ. \textbf{50}, 102116 (2025)

\bibitem{retOvgun:2018ran} A.~Ovg{\"u}n, K.~Jusufi and I.~Sakalli, Gravitational lensing under the effect of Weyl and bumblebee gravities: Applications of Gauss{\textendash}Bonnet theorem. Annals Phys. \textbf{399}, 193 (2018)

\bibitem{Hassanabadi:2025ect} H.~Hassanabadi, A.~Guvendi, F.~Kafikang, T.~Sathiyaraj and S.~Zare, Ergosphere Dynamics and Rotational Energy Extraction in Bumblebee Kerr-Newman-AdS Black Holes. [arXiv:2512.18512 [gr-qc]]


\bibitem{retDeng:2025uvp} W.~Deng, W.~Liu, F.~Long, K.~Xiao and J.~Jing, Quasinormal modes of a massive scalar field in slowly rotating Einstein-Bumblebee black holes. JCAP \textbf{11}, 028 (2025)



\bibitem{retMattingly} D.~Mattingly, Modern tests of Lorentz invariance. Living Rev. Rel. \textbf{8}, 5 (2005)

 \bibitem{retFilho:2021rin} A.~A.~A.~Filho and A.~Y.~Petrov, Higher-derivative Lorentz-breaking dispersion relations: a thermal description. Eur. Phys. J. C \textbf{81}, 843 (2021)
\bibitem{retAnacleto:2018wlj} M.~A.~Anacleto, F.~A.~Brito, E.~Maciel, A.~Mohammadi, E.~Passos, W.~O.~Santos and J.~R.~L.~Santos, Lorentz-violating dimension-five operator contribution to the black body radiation. Phys. Lett. B \textbf{785}, 191 (2018)
\bibitem{retFilho:2020cgk} A.~A.~A.~Filho, Lorentz-violating scenarios in a thermal reservoir. Eur. Phys. J. Plus \textbf{136}, 417 (2021)
\bibitem{retFilho:2023ydb} A.~A.~A.~Filho, J.~Furtado, H.~Hassanabadi and J.~A.~A.~S.~Reis, Thermal analysis of photon-like particles in rainbow gravity. Phys. Dark Univ. \textbf{42}, 101310 (2023)



\bibitem{retShi:2025rfq} Y.~Shi and A.~A.~Ara{\'u}jo Filho, Influence of a Kalb-Ramond black hole on neutrino behavior. JHEP \textbf{08}, 028 (2025)

\bibitem{retShi:2025plr} Y.~Shi and A.~A.~Ara{\'u}jo Filho, Effects of bumblebee gravity on neutrino motion.
JCAP \textbf{11}, 045 (2025)

\bibitem{retBarriola:1989hx} M.~Barriola and A.~Vilenkin, Gravitational Field of a Global Monopole.
Phys. Rev. Lett. \textbf{63}, 341 (1989)

\bibitem{retVilenkin:1981kz} A.~Vilenkin, Cosmic Strings, Phys. Rev. D \textbf{24}, 2082-2089 (1981)

\bibitem{retVilenkin:1984ib} A.~Vilenkin, Cosmic Strings and Domain Walls. Phys. Rept. \textbf{121}, 263-315 (1985)



\bibitem{retNooriGashti:2025vyg} S.~Noori Gashti, Y.~Sekhmani, M.~A.~S.~Afshar, M.~R.~Alipour, M.~Khodajou Masouleh, B.~Pourhassan, J.~Sadeghi and J.~Rayimbaev, Thermodynamic signatures in black hole geometry and harmonic oscillations with nonlinear electromagnetic fields and phantom global monopole. Nucl. Phys. B \textbf{1022}, 117244  (2026)

\bibitem{retAhmed:2025yyb} F.~Ahmed and A.~Bouzenada, Ray geodesic and wave propagation in lorentz-violating wormholes with topological defects. Phys. Scripta \textbf{100}, 125306 (2025)

\bibitem{retYousaf:2025jqx} M.~Yousaf, H.~Asad, K.~A.~Yasir, S.~K.~Maurya, A.~Ali and F.~Atamurotov, Global monopole induced wormholes in power-law gravity: stability and physical viability. Eur. Phys. J. C \textbf{85}, 1352  (2025)

\bibitem{retSekhmani:2025yjf} Y.~Sekhmani, S.~K.~Maurya, J.~Rayimbaev, A.~Ali, M.~K.~Jasim, I.~Ibragimov and S.~Muminov, Black holes surrounded by a Murnaghan-fluid scalar gas in a global-monopole background.
Phys. Dark Univ. \textbf{50}, 102176  (2025)




\bibitem{retDadhich} N. Dadhich, K. Narayan, U.A. Yajnik, Schwarzschild black hole with global monopole charge. Pramana-J. Phys. \textbf{50}, 307-314 (1998)
\bibitem{retGullu:2020qzu} {\.I}.~G{\"u}ll{\"u} and A.~{\"O}vg{\"u}n, Schwarzschild-like black hole with a topological defect in bumblebee gravity. Annals Phys. \textbf{436}, 168721 (2022)

\bibitem{retMaki} Z.~Maki, M.~Nakagawa and S.~Sakata, Remarks on the unified model of elementary particles, Prog. Theor. Phys. \textbf{28} 870 (1962)

\bibitem{retSwami}H.~Swami, Neutrino flavor oscillations in a rotating spacetime, Eur. Phys. J. C \textbf{82} 974 (2022)

\bibitem{retBilenky:1978nj}S.~M.~Bilenky and B.~Pontecorvo, Lepton Mixing and Neutrino Oscillations. Phys. Rept. \textbf{41}, 225 (1978)

\bibitem{retChakrabarty:2023kld} H.~Chakrabarty, A.~Chatrabhuti, D.~Malafarina, B.~Silasan and T.~Tangphati, Effects of gravitational lensing by Kaluza-Klein black holes on neutrino oscillations. JCAP \textbf{08} 018 (2023)



\bibitem{retEsteban:2018azc} I.~Esteban, M.~C.~Gonzalez-Garcia, A.~Hernandez-Cabezudo, M.~Maltoni and T.~Schwetz, Global analysis of three-flavour neutrino oscillations: synergies and tensions in the determination of $\theta_{23}$, $\delta_{CP}$, and the mass ordering. JHEP \textbf{01}, 106 (2019)

\bibitem{retVilenkin} A. Vilenkin, E.P.S. Shellard, Cosmic Strings and Other Topological Defects, Cambridge University Press, Cambridge, 1994.

\bibitem{retPantig:2024kqy} R.~C.~Pantig, On the analytic generalization of particle deflection in the weak field regime and shadow size in light of EHT constraints for Schwarzschild-like black hole solutions, Eur. Phys. J. C \textbf{85}, 52 (2025)


  \bibitem{retWu:2022eiv} S.~R.~Wu, B.~Q.~Wang, D.~Liu and Z.~W.~Long, Echoes of charged black-bounce spacetimes. Eur. Phys. J. C \textbf{82},  998 (2022)

\bibitem{retWu:2024hxr} S.~R.~Wu, B.~Q.~Wang, Z.~W.~Long and H.~Chen, Rotating black holes surrounded by a dark matter halo in the galactic center of M87 and Sgr A{\ensuremath{*}}. Phys. Dark Univ. \textbf{44}, 101455 (2024)

\end{thebibliography}
\end{document}